# Observing lightning and transient luminous events from the International Space Station during ILAN-ES: an astronaut's perspective

Yoav Yair*(1), Melody Korman (2,3), Colin Price (2) and Eytan Stibbe (3)

(1) School of Sustainability, Reichman University (IDC Herzliya), Israel
(2) Porter School of the Environment and Earth Sciences, Tel Aviv University, Israel
(3) Rakia Mission, Tel-Aviv, Israel

**\*Corresponding author**

Prof. Yoav Yair

School of Sustainability, Reichman University (IDC) Herzliya

P.O. Box 167, 8 University Street, Herzliya 4610101 Israel

(p) +972-9-9527952 (m) +972-52-5415091 (f) +972-9-9602401

Email: yoav.yair@runi.ac.il

Keywords: Lightning; Thunderstorms; Transient Luminous Events; International Space Station; Earth Observations

Abstract

The ILAN-ES (Imaging of Lightning And Nocturnal Emissions from Space) experiment was conducted by Israeli astronaut Eytan Stibbe in April 2022 as part of the Axiom Space company AX-1 private mission to the International Space Station, in the framework of Rakia, a set of experiments selected for flight by the Ramon Foundation and the Israeli Space Agency. The mission objective was to manually record lightning and transient luminous events from the Copula window in the ISS, based on preliminary thunderstorm forecasts uploaded to the crew 24-36 hours in advance. A Nikon D6 camera with a 50 mm lens was used, in a video mode of 60 frames per second. During the 15-day mission, 82 different targets were uploaded to the ISS of which 20 were imaged by the astronauts, yielding a total harvest of 45 TLEs: sprites, Elves and Blue Corona Discharges. The methodology and execution by the ISS astronauts are described and recommendation for future observations by on-board human-operated instruments on the ISS are given.

1. Introduction

Sprites, haloes, and ELVES seem to be ubiquitous in the mesosphere above thunderstorms around the planet and were reported over most major centers of lightning activity on Earth. However, their global rate is still elusive and was not unequivocally determined (Chern et al., 2008). Thus, observations of the Earth from space offer an advantageous vantage point which allows the continuous monitoring and large-scale coverage of these atmospheric phenomena. A space based observing platform of lightning-induced TLEs can overcome many limitations of ground-based measurements, especially the inherent locality of a site-based measurement and the line-of-site and cloud obscuration issues. The global rate of occurrence can be deduced and correlated with the meteorological characteristics of the parent storms.

Astronaut-controlled directed observations of lightning and TLEs from space were first conducted during the Mediterranean Israeli Dust Experiment (MEIDEX) on board the Space Shuttle Columbia in January 2003 (Yair et al., 2003). During that 16-day mission, the primary daytime mission was to study dust effects on clouds and on radiative transfer over the Mediterranean Sea; as a secondary science objective, global imaging of TLEs from space was added for the nocturnal part of the orbit. Due to the high orbital speed of 8 km s$^{-1}$, the potential thunderstorm targets had to be defined in advance at least 24 hours prior to the allocated observation slot. The observation was limb-pointing, such that a vertical separation between the parent lightning and the induced TLE would enable a clear separation in the image (Boeck et al., 1998). During the mission the ground crew in the Payload Operations and Control Centre (POCC) at NASA GSFC used lightning locations derived from the time of group arrival (TOGA) network (later to become the World Wide Lightning Location Network, WWLLN; Rodger et al., 2006) to improve on the initial forecast. This enabled a 4.5-hour closure before the actual observation time and fine-tuning of the camera pointing angle, which was relayed through mission control to the crewmember who was conducting the observation. As reported by Yair et al. (2004), out of 24 orbital windows allocated for TLE observations during the mission, almost 6 hours of valuable data were saved (see their Table 1 for details) with 17 TLEs detected - a significantly higher detection rate compared with fixed-camera campaigns from the shuttle (the Mesoscale Lightning Experiment, Boeck et al., 1998) or from the ISS (the LSO, performed by French and Belgian astronauts; Blanc et al., 2004).

In 2011, JAXA astronaut Dr. Satoshi Furukawa conducted the "Cosmic Shore" campaign, as part of Expedition 28/29 to the ISS. The observation profile was more flexible in that it enabled pointing the camera through the Cupola window, based on the astronaut's view of the atmosphere below. For maximizing the limited available observation time, Yair et al. (2013) used the same methodology of using SIGWX maps to give a 48-hours advanced notice to the crew, integrated into the daily JEDI message uploaded to the ISS. Unlike the space shuttle, the ISS cannot be maneuvered to offer a better line-of-sight to a potential thunderstorm target, but this deficiency is more than compensated for by the large field-of-view offered by the Copula. The "Cosmic Shore" used a color EMCCD TV camera, and succeeded in documenting 10 TLEs in true color, along with first image of a sprite-halo and the first unambiguous nadir-view image of a sprite, horizontally displaced from the parent flash (Yair et al., 2013). The shortcoming of that campaign was the lack of an accurate time stamp on the camera image (~1 second) and no reliable recording of the pointing azimuth, a fact that limited the capability to correlate the observed TLE with lightning locations detected by global networks such as the WWLLN. Nevertheless, the campaign highlighted once more how astronaut-time can be utilized efficiently on-board the ISS when accurate lightning forecasts are given ahead of time. Another directed TLE campaign from space was the THOR experiment that was conducted by the Danish astronaut Andreas Mogensen during his short duration mission (IRISS) to the ISS during 1-10 September 2015. The aim was to observe over-shooting cumulonimbus turrets, lightning and high-altitude TLEs with an optical camera through the windows of the space station. To support the experiment, we developed a strategy to predict locations of thunderstorm targets up to three days in advance, the long lead time required by the astronaut activity planners to accommodate many experiments during short time missions. New and additional components were added to the forecast that enabled us to distill and prioritize a daily target-list with specific viewing angles computed relative to the ISS position and altitude. The results of the THOR campaigns and the discovery of new electrical phenomena were published by Chanrion et al. (2017).

In parallel with occasional crew-controlled observations, several autonomous payloads on-board the ISS had also succeeded in detecting TLEs from orbit during long-duration operations. The Japanese JEM-GLIMS experiment operated from 2011 through 2015 (Sato et al., 2016). During nearly 4 years of operation GLIMS detected 8537 combined events of superimposed lightning and TLEs, and post-analysis differentiated 699 TLEs, of which 508 were Elves and 42 sprites, with 149 events yet to be determined. Neubert et al. (2021) reported the successful imaging of the development of a blue jet emerging from the top of the thundercloud, achieved by the MMIA which is part of the ASIM (Atmosphere-Space Interaction Monitor; Chanrion et al., 2019; Neubert et al., 2019) payload on the ISS, through automatic recording with no prescribed manual operation by a crew member or from the ground. Additional results of BLUE events and high-energy emissions were reported by Bjørge-Engeland et al. (2020), Liu et al. (2021) and Skeie et al. (2022) and are reviewed by Neubert et al. (2023). These two payloads demonstrate clearly that a camera on orbit can detect TLEs and other lightning-related emissions quite frequently.

2. The ILAN-ES mission design

The ILAN-ES campaign was planned as part of the Rakia mission, in the framework of the first private expedition to the ISS by the Axiom Space company (AX-1). It was selected by The Rakia Scientific Committee, comprised of representatives from Israeli universities, the Ramon Foundation and the Israeli Space Agency. It that mapped more than 70 space-borne experiment proposals by Israeli start-up companies and scientists. The ILAN-ES was one of the 34 experiments selected for flight and was designed as a heritage of the MEIDEX campaign conducted in 2003, with an intention to optimize the observation time from the ISS by defining only selected targets at prescribed times to be imaged by the astronaut. This was planned to guarantee a high scientific yield from a rather short mission. The added value of combined ground-based observations of the parent lightning, and the potential for simultaneous optical imaging from the ground of specific events were expected to produce new scientific data on the 3-dimensional structure of TLEs. The mission was launched on a Space-X Falcon 9 launcher on April 8th, 2022, and returned to earth on April 25th, 2022, completing more than 17 days in orbit. Israeli astronaut Eytan Stibbe was one of the crew of 4 and was assigned the title PA-3 (Private Astronaut 3) in the mission plan. The following sections describe various aspects of performing observations from the Copula window by the astronaut.

2.1 Thunderstorm targets in the ISS crew timeline

In order to prepare the crew in advance for the times (=geographical locations) of potential thunderstorm targets, Ziv et al. (2004) devised a methodology that relied on the Significant Weather Maps (SIGWX) issued daily by the Aviation Weather Centre (AWC). During the AX-1 mission, these maps were automatically retrieved by the Israeli Meteorological Service and uploaded to an FTP server. They highlight areas of deep convection and indicate the potential locations for embedded/isolated cumulonimbus clouds, the percentage of coverage and their anticipated cloud-top heights. An example of a map used for the prediction of Cumulonimbus cloud-top heights at T+36 h from 19 April 2022 is given in Figure 1. Based on Cumulonimbus top heights predicted to be above ~14 km and other meteorological considerations (Ziv et al., 2014; Yair et al., 2016) several potential targets were defined. Considering the predicted ISS orbit (using the Satellite ToolKit [STK] software), a list of targets was prepared and transmitted daily to NASA/TCO team, who injected them in the daily Crew Earth Observations (CEO) slot of the Execute Package [or JEDI] for 22 April 2022 (Appendix 1).

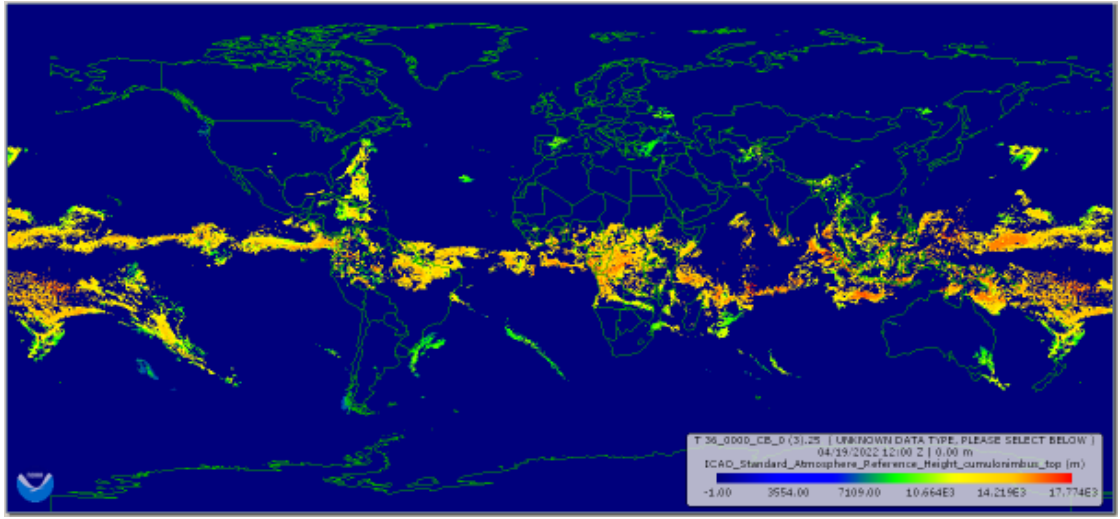

Figure 1: Forecast map of Cb height (in km) from 19 April 2022 12:00 UT valid for 36 hours.

The CEO mode of operation has been in practice for many years on the ISS, with considerable scientific yield (Gebelein and Eppler, 2006; Robinson et al., 2002). Thus, mission planners were able to define viewing slots to be executed by the astronaut at specific times (15-20 minutes each) for the overflight of the chosen region of interest (ROI). The selected targets were put in the ISS OpTIMIS (Operations Planning Timeline Integration System; Beard, 2018) viewer of flight information that displays on an iPad the daily activities for the entire crew in a graphic display that also contains night/day and TDRS satellite coverage availability. Time-sensitive experiments or tasks are marked in blue in each astronaut's daily assignments list (Figure 2).

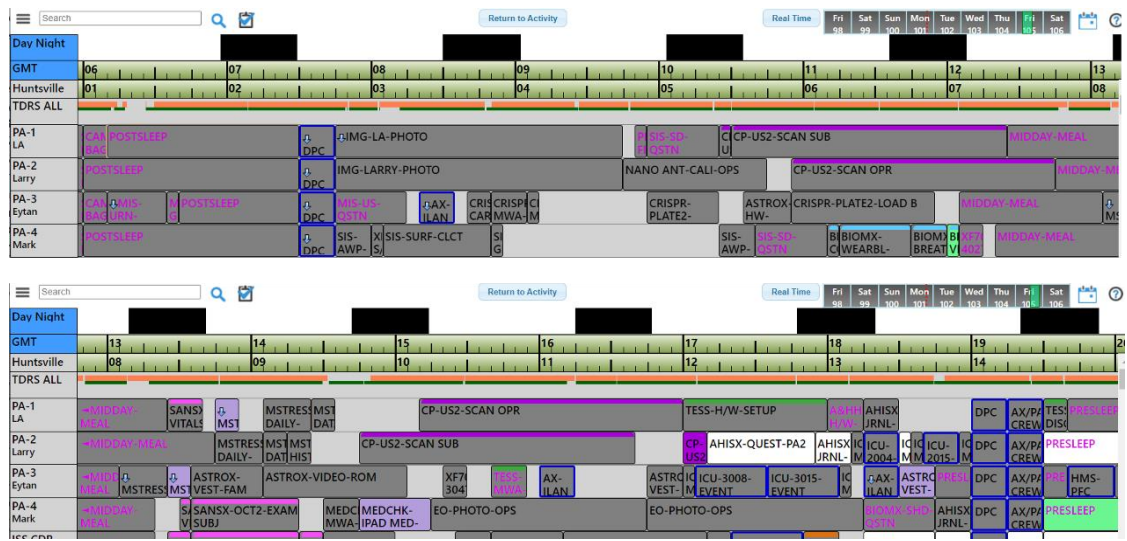

Figure 2: Excerpt of OpTIMIS timeline for Day 7 (15 April 2022) for the AX-1 crew. Note that PA-3 has 3 ILAN-ES targets, marked in blue.

Since the standard definition of working hours for astronauts on the ISS are from 0700-2000 UT, the targets had to be sent no later than 1100 UT the day before the actual observation, a fact that also limited the list of potential targets to the specific nocturnal segments of orbits traversed by the ISS during the crew's working hours. The reality of the crew availability during the Rakia mission dictated a prioritization of the available targets, such that the science team had to select up to 5 daily observation windows where the probability of intense lightning activity was evaluated to be highest. These targets were relayed to the crew as high-priority tasks, even though the reality evolving during mission operation had the astronauts make judgements about their ability to make a full observation run at the specified time. Every so often, due to conflicts in scheduling and delays incurred by previous astronaut tasks, the ILAN-ES targets were skipped.

2.1 Camera set-up and operation in the Copula

The camera that was used in the ILAN-ES campaign was a Nikon D6 set at 6400 ISO and recording 60 frames per second at 1920 x 1080 pixels. It was equipped with a 50mm/f1.2 lens, giving a 34.4º x 19.75º field of view corresponding to 1.07'/pixel. Focus should be set at infinity and verified. With these settings, the camera resolution was 130 m at nadir. For ILAN-ES, the camera was brought on-board the ISS by the AX-1 crew and is not (as of yet) a part of the existing inventory of ISS cameras. The astronaut operated the camera from the Copula window of the ISS while visually tracking lightning activity and directing the camera towards the bright flashes, such that they will occupy the lower part of the frame, leaving a part of the sky to occupy upper atmospheric discharges. This allowed capturing TLEs occurring above the visible flashes during slant or limb views, which offers better spatial separation between the lightning and the TLE above it, and a greater probability of a successful observation. Each observation run was scheduled to be 5-10 minutes long, as targets span a few hundreds of km in size.

The time that it takes to set-up the camera for the proper imaging configuration may take up to 10 minutes, as cameras in the Copula are not always located at given fixed locations and may be absent altogether due to previous usage, and so the total time block that should be defined for a CEO slot should be of the order of 20-25 minutes. As for the getting organized inside the Copula, it is required that for the imaging sequence the camera will be wrapped in a manner that diminishes internal reflections, and an appropriate hood made of black and opaque material is placed around the camera body and lens. Care is taken during the observation that the camera does not to touch any of the windows, to avoid internal scratches and permanent damage. Also, lighting from other parts of the ISS close to the Copula should be minimized as well in order to avoid stray-light contamination.

2.2 Pointing and tracking

Floating in microgravity makes target acquisition and tracking a challenging task, as the pointing towards the target relies on visual identification of the thunderstorms. The view from the ISS enables identifying lightning at ranges from nadir-to-limb, a projected area with a radius of ~2300 km (depending on the ISS height at the specific part of the orbit). When the defined time of the observation approaches, the large horizontal dimension of the storms enables early visual identification, provided that the astronaut knows to which of the Copula windows to focus attention to. Normally the definition of port/starboard is based on the ISS velocity-vector frame-of-reference. However, the astronaut in the Copula floats upside down with head towards the earth below the ISS, a fact that may create disorientation as to which side is port and which is starboard in the astronauts' frame-of-reference. This may be exacerbated by the lack of motion cues during orbital night, which may make it harder to properly determine the direction of the ISS motion against the dark background. Note that the camera FOV is limited compared with the unobstructed view through the Copula window, which enables identifying the lightning targets rather easily. And so, for recording the data with the camera, viewing through the camera eyepiece is not practical and the scenery is best viewed by looking at the camera screen, although admittedly this too offers a limited area determined by the angle of pointing.

Thunderstorms may occupy large areas, when embedded in Mesoscale Convective Systems (MCS) or Mesoscale Convective Complexes (MCC) that can have areas larger than 20,000 $km^2$ (Goodman and MacGorman, 1986; Houze, 2004). As the ISS traverses the predicted area at 8 km $s^{-1}$ the total time available to the astronauts is, on average, 5-7 minutes, depending on the position of the MCC with respect the ISS track. The astronaut can extend the time of the observations by dynamically changing the pointing angle of the camera, tracking the main electrical activity centers. Often this may necessitates switching windows within the Copula during the observation run, which requires agility and stable tracking while manually holding the camera and directing it to the slowly moving target. For post flight analysis, it is important to have pointing information registered as part of the Crew Notes on the individual iPad and recorded. If missing, the pointing direction can be reconstructed by noting (if present) visible parts of the ISS within the image, for example the Soyuz capsule or the solar panels.

2.3 Time information

A crucial element that is essential for scientific observations from space is the timestamp on the video image. This is especially crucial when searching from sub-millisecond transient luminous events which are generated by lightning. Since all the above-mentioned ground-based lightning detection systems use GPS time source for event registrations, a basic mission requirement is that the time stamp on the camera image is as accurate as possible. The time keeping in the ISS is maintained through a combination of atomic clocks and time signals from ground-based radio stations, allowing for periodic updates from mission control. Thus, when operating the camera, it is one of the astronauts' paramount duties to ensure that the camera time is as close as possible to ISS time throughout the mission.

Analyzing the data obtained during ILA-ES shows that there was an offset between the time on the images and the GPS time base of lightning detection systems. This is best exemplified in the sprite event observed on 15 April 2022 at 08:32:51.48 UT lasting 2 frames, following a lighting that was first detected at 08:32:50.86. The target was predicted 36 hours in advance as an oceanic convective center west of Nouméa, New Caledonia, and the start time of the observation window was defined at 08:27:00 UT (ground location 22.27 S 166.44 E) for 5 minutes, the closest point at nadir viewing being defined at 26.91 S 152.63 E. This was the 3rd target out of 7 defined for that mission day, of which only 2 were executed. Figure 3a shows the frame containing the sprite, which was frame number 38 out of 79 frames of the parent lightning luminosity (each frame is 1/60 seconds or 0.01667 s). The 0.6334 s delay between lightning onset and sprite appearance falls within the category of "long delayed" sprites (Li et al., 2008; Soula et al., 2017). The total time of lightning light viewed by the camera was 1.316 seconds, indicating a long continuing current, a component typical of sprite-producing flashes (Bell et al., 1998 Cummer and Stanley, 1999).

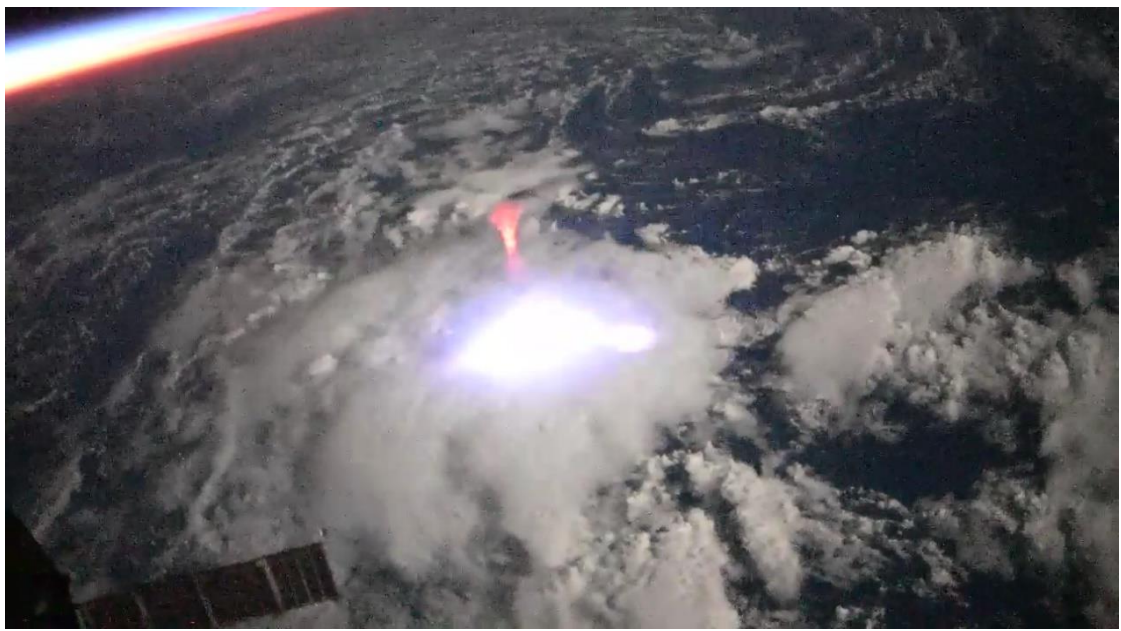

Figure 3a: sprite above a Cb anvil south-west of French Caledonia, 15 April 2022 08:32:54.48 UT. The sunset near the limb, viewing direction north-west. The range of the lightning from the ISS was 682 km. Image by Eytan Stibbe/ Rakia Mission/ ILAN-ES Science Team

Based on ENTLN data, a +128.72-kA flash that occurred at 08:32:54.700 was the most likely candidate for the sprite parent. It was located at 28.3123 S 162.971 E and was observed obliquely as the ISS was travelling south-east, from a range of 682 km. The WWLLN identified this flash at 08:32:54.701 at the exact same coordinates. The other flashes in that storm before and after the sprite were either intracloud or weak negative CGs, incapable of generating a sprite (see table in the Appendix 2). Based on the time of the sprite within the continuous current there is an offset of 3.84 seconds between the camera time and the GPS time-base of the ground systems, the camera being ahead (08:32:54.70 – 08:32:50.86). This offset has little significance in terms of ISS location at the time of the event (~30 km) but is crucial for the timing of the parent flashes for other TLE, specifically 2 Elves that were observed above the same Cb cell at 08:35:28.4 and 08:36:01.29 UT (Figure 3b). It can be concluded that the insertion of the time on the Nikon D6 should be conducted by the astronaut as accurately as possible relative to the ISS time and verified for any drift along the duration of the mission. Post-analysis, by carefully matching optical and ground-based observations of specific events (lightning or TLEs or both), the precise offset between GPS and ISS camera time can be deduced, and necessary corrections can then be implemented for the entire database.

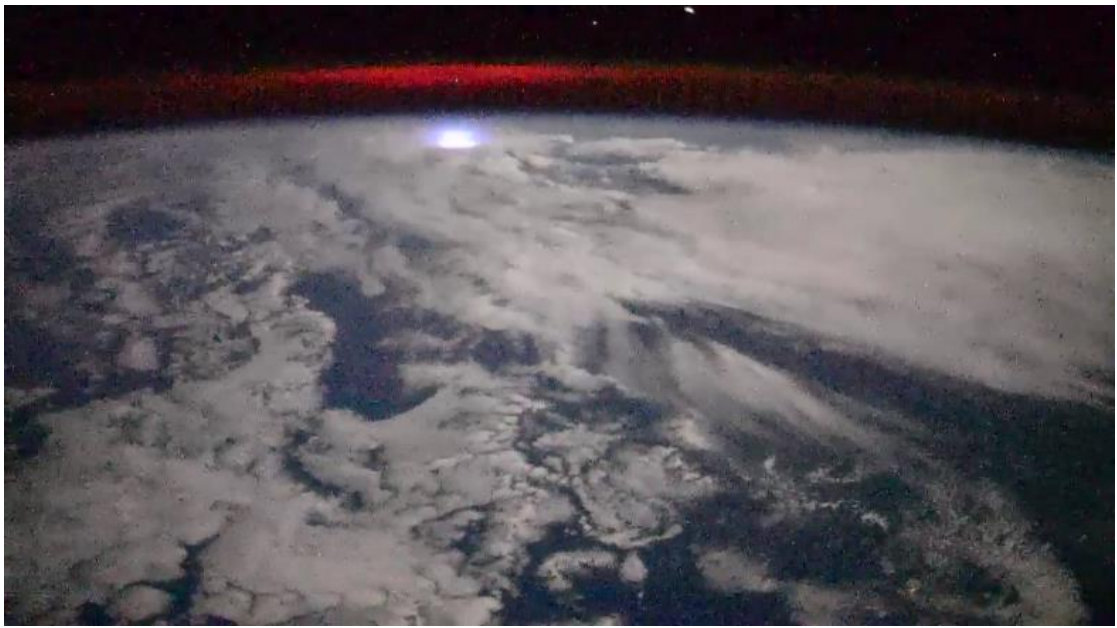

Figure 3b: Elves the same thunderstorm cell, 15 April 2022 08:35:28.40 UT. The range of the parent flash from the ISS was 1813 km. Image by Eytan Stibbe/Rakia Mission/ILAN-ES Science Team

2.3 Data handling and crew notes

After completing the observations and evacuating the Copula, the memory card of the camera is taken by the astronaut and copied to a free ISS workstation for downloading the video. The files are uploaded into a Download subdirectory and are given names according to the experiment they are related to (or any convention agreed upon) and sent to the ground for processing and distribution. It may take up to 48 hours for the data to be available for analysis via a Box file-share application. After the data had been transferred, the camera card is cleared for new usage. Thus, during the mission, the science team has a delay before they can evaluate the quality of the data obtained during the observations. This may be of great importance if changes are needed, for example of the lens or the camera setting. During ILAN-ES there was a fear of high noise level in the images, and an attempt to decrease the ISO level to 3200 was made (instead of the nominal value 6400). Such a change was related to the crew by the TCO and implemented in several observation runs, and upon analysis the setup was restored to the original configuration.

Astronauts can add succinct and short Crew Notes which constitute a textual description of the observed scene. During ILAN-ES the information describing the intensity of the storm and the observed flash-rate (high, low, none) and any outstanding phenomena ("Blue Flash"?). These crew notes were available to the science team through the representative in Mission Control with access to OpTIMIS only after completing the observation, long before the data was downloaded.

3. Short duration observations

Observations conducted by astronauts at night require agility, focus and flexibility in changing spatial orientation during the execution. Although they may last only 5-8 minutes, there still exists a potential for high scientific yield from even shorter observations, of 1 minute or less. To exemplify this, we review an observation of a thunderstorm that occurred on 21 April 2022, included in the last orbit that was in the target list (Appendix 1), where the astronaut was instructed to conduct nadir pointing beginning near Madagascar at 21:17:00 UT while the ISS was moving above the Indian Ocean north-east towards Myanmar, with sunrise expected at the end of the observation. The observations began at (corrected time for offset, as explained above) 21:12:22.84 UT and was divided into 9 separate very short video segments. The last video segment began at 21:30:27.84 UT as the ISS was above the Andaman Sea and was just 30.33 seconds long. Moreover, the last 13 seconds were already disrupted by sunlight as the ISS was approaching the day terminator, leaving just ~17 seconds of useful data.

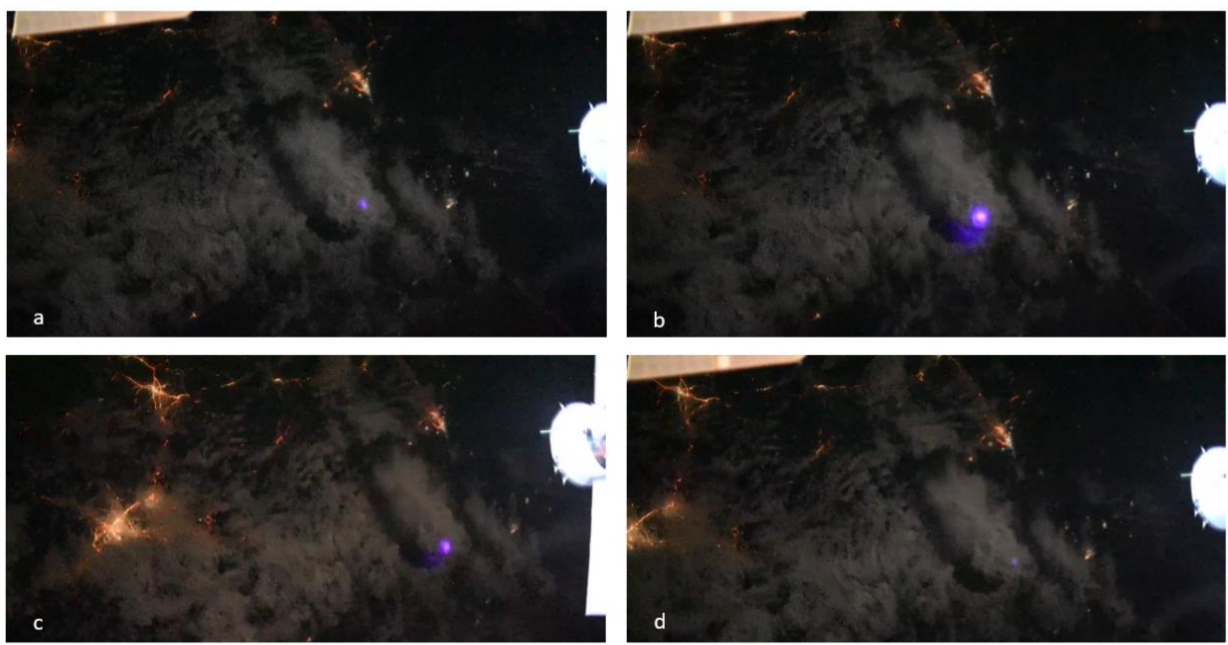

Figure 4: Nadir view of 4 blue corona discharges detected during a 17-second observation run above Port Blair, India, 21 April 2022. City light are clearly visible beneath the cloud cover. The selected events occurred at (a) 0.57 s (b) 2.22 s (c) 4.87 s (d) 14.11 s after 21:30:27.84 UT. Sunrise for the ISS occurred at 17.33 s. Image by Eytan Stibbe/Rakia Mission/ILAN-ES Science Team

The astronaut viewed a penetrating turret of an active Cumulonimbus cell located above the city of Port Blair, India (11°40' N 92°44' E). At 21:28:00 the ISS was located at 11.47 N 92.77 E, slightly to the east of the city. In the span of 8.5 seconds, 11 Blue Corona Discharges (Dimitriadou et al., 2022) events, each lasting 1 frame, were recorded over the anvil and the penetrating turret of the Cumulonimbus, intermittent with the lightning produced by this cell. Figure 4 shows some of these BLUEs in unprecedent clarity. These blue emissions are accompanying Narrow-Bipolar Events (NBEs) and emit mostly in 337 nm, indicating that they are corona discharges (Liu et al., 2021). All the blue events observed here appeared to be located on the top of the cloud, a fact that indicates that they are occurring between the upper positive charge center and the negative screening layer just above the cloud top (see Figure 4 in Neubert et al., 2023). The computed rate is 0.64 events per second, or an average occurrence of a blue event every ~1.54 seconds, in agreement with the rate reported by Chanrion et al. (2015). (Further analysis of the meteorological and electrical characteristics of that storm are beyond the scope of this manuscript and will be published in a forthcoming paper). This example demonstrates the high value of even very short video recordings of active thunderstorms and exemplifies the added accuracy in target selection in real time by the human-controlled camera pointing through the Copula window.

4. Ground campaign

As part of the science objectives of the mission, the ILAN-ES was augmented by a wide network of ground-based observations, to attempt simultaneous space-ground TLE and lightning observations. There are a couple of citizen-scientists TLE detection networks conducting optical observations by ground-based cameras operating in central Europe (centered in Hungary and Croatia), the Caribbean (Puerto Rico) and south America. Although the probability of simultaneous observations from space and ground is small, the global coverage offered by the various amateur networks enhances the chance for space-ground detection. See: https://www.facebook.com/groups/376355972487572
Additionally, we established a collaboration with international teams of researchers from universities. These included (1) University of Toulouse facilities on the Pic du-Midi observatory (Pyrenees mountains) and the Isle de Reunion (Indian Ocean); (2) UPC Barcelona, and sites in Santa-Marta, Colombia; (3) The camera network of the Chinese Academy of Sciences (Beijing) and (4) the LEONA network in central Brazil. As an educational component that had a priority for the entire Rakia mission, we distributed cameras to schools in Israel, Ghana, Rwanda, Zimbabwe, Brazil and Hong-Kong, with the hope of getting high school students involved in optical observations in conjunction with ISS overpasses, based on local thunderstorm forecasts.

During the mission, a team of students from the School of Sustainability at Reichman University oversaw communication with the schoolteachers, informing them on the over-flight times of the ISS and encouraging observations of thunderstorms, if they were indeed forecasted in the area of the school. Unfortunately, most of the schools went on Easter holiday during the mission and were unable to conduct observations in conjunction with the ILAN-ES overflight. As for the other ground locations, while there were some good opportunities for a mutual observation of thunderstorms, they were not successful, mostly because of cloud obscuration at the ground station or technical malfunctions (S. Soula, O. van-der Velde, personal communication). This does not diminish the importance of attempting future ground-space-based coordinated observations, and most importantly, offers valuable opportunities for STEM education (Bonnie et al., 2014; Phillips et al., 2019). The recently launched NASA citizen science project "Spritacular" may become such a platform (https://spritacular.org/; Accessed 1 March 2023).

5. Conclusions

The ISS is an orbiting platform that offers an unparalleled vantage point for global atmospheric observations. There are more than a dozen cameras onboard, and they are used by astronauts for various tasks and Earth Observations, but mostly in a target-of-opportunity mode or on-demand for specific Earth-events such as tropical storms, volcanic eruptions and forest fires, which are then shared with the general public via NASA's Earth Observatory (https://earthobservatory.nasa.gov/; Accessed 2 March 2023). Most of the imaging is done during daytime after specific times are inserted into the crew' daily execute package in the Crew Earth Observation message format. However, very few, if any, targets are defined for nocturnal observations, and so for most of orbital night the Copula is an underused asset.

Thunderstorms are frequent almost everywhere on earth, especially over land in the tropics but also above the oceans and are easily observable from space. Future scenarios of climate change predict an increase in the prevalence, intensity and geographical distribution of thunderstorms and lightning with significant climatic implications (Chen et al., 2021; Qie et al., 2021). Daily targets over the main convective regions of earth can be easily defined, based on the meteorological features of sprite-producing storms in the summer and winter hemispheres (Williams and Yair, 1006).

The continued operation of the ASIM payload on the ISS (Neubert et al., 2023) and the addition of more new observational capabilities such as the Falco-Neuro payload of the US Air-Force (McHarg et al., 2022) make human-directed optical observations very valuable. Achieving simultaneous observations by 2 or 3 platforms on the ISS of such lightning-related phenomena is not complicated and can be achieved with careful planning. The operational methodology and scientific value of human directed observations that were described above support continued research of thunderstorms and TLEs from space. This type of research entails little additional budget and can be a complementary science objective for crews in future private space missions. New long-duration missions will likely enhance imaging capabilities, by bringing high framerate and high-resolution cameras on board the ISS. The need for optimizing crew time can be implemented based on the experience gained during the Rakia mission and the ILAN-ES experiment.

**Acknowledgement**

The ILAN-ES project was supported by Reichman University Internal Research Fund and by the Canadian Friends of Tel-Aviv University. We thank Denmark Technical University for allowing calculations of ISS locations through their ASDC server.

We thank Nir Stav, Evgeny Brainin and Vladimir Meerson from the Israeli Meteorological Service for technical support in obtaining Aviation Weather maps.

We wish to thank Eliran Hemo (Ramon Foundation), Brandon Williams (Axiom Space), Jim Watson, Ryan Miller and the TCO team (NASA-MSFC) for their help in mission planning and operation.

The student team from the School of Sustainability at Reichman University included Noah Gordis, Adam Bernitz, Linoy Levi, Bar Amrami, Bosco Nshimiyumukiza, Yuval Sheleg, Ryan Khalill, Daniel Telkar and Rebecca Mingeleva; Administrative support by Tali Livne.

**Appendix 1**

Table showing the list of targets transmitted to NASA/TCO through the Axiom mission support for 21 April 2022. Out of the 14 targets, only 5 were prioritized for observation (marked in yellow)

| Daily Orbit Number | Central reference Location | Central reference point | Start Time GMT | Start point | Closest Time GMT | Closest point | Additional info |
|---|---|---|---|---|---|---|---|
| 5 | Cusco, Peru -> Colombia | 13.53 S 71.96 W | 07:24:00 | 12.79 S 72.37 W | 07:30:00 | 5.49 N 50.25 W | Thunderstorms Port |
| 6 | Auckland New Zealand | 36.85 S 174.76 E | 08:40:00 | 51 S 157 E | 08:45:00 | 44 S 132 E | Thunderstorms Starboard |
| 6 | Galapagos -> Panama | 0.18 S 78.47 W | 09:00:00 | 3.1 S 88.94 W | 09:04:00 | 9.06 N 80.27 W | Thunderstorms Nadir & Port |
| 7 | Auckland New Zealand | 36.85 S 174.76 E | 10:10:00 | 51.73 S 161.82 E | 10:15:00 | 48.77 S 169.4 W | Thunderstorms Port |
| 8 | Auckland New Zealand | 36.85 S 174.76 E | 11:50:00 | 45 S 177 E | 11:55:00 | 34 S 163 E | Thunderstorms Port |
| 10 | Townsville Australia -> King Solomon | 19.25 S 146.8 E | 15:05:00 | 22.0 S 162.1 E | 15:12:00 | 1.04 S 178.1 E | Thunderstorms Port |
| 11 | Surabaya, Indonesia | 7.25 S 112.75 E | 16:40:00 | 15.69 S 143.8 E | 16:45:00 | 0.05 N 154.9 E | Thunderstorms Nadir & Port |
| 12 | Jakarta Indonesia | 6.2 S 106.8 E | 18:15:00 | 9.1 S 125.2 E | 18:22:00 | 4.89 N 105.4 E | Thunderstorms Nadir & Port |
| 13 | Banda Aceh Indonesia | 5.54 N 95.32 E | 19:50:00 | 2.5 S 106.36 E | 19:54:00 | 9.68 N 115.0 E | Thunderstorms Starboard |
| 14 | Madagascar | 16.17 S 49.76 E | 21:17:00 | 20 S 69.5 E | 21:21:00 | 8.0 S 78.7 E | Thunderstorms Nadir & Port |

# Appendix 2

Table showing ENTLN data of lightning flashes detected on 15 April 2022 in conjunction with the sprite shown in Figure 2A. The yellow-highlighted positive cloud-to-ground flash is the most likely sprite-parent.

| UTC | lat | lon | type | peak_current | peak_current_KA | ic_height |
|---|---|---|---|---|---|---|
| 08:32:47.204 | -21.0514 | 170.1377 | 0 | 77798 | 77.798 | 0 |
| 08:32:47.268 | -21.1155 | 169.9281 | 0 | 21332 | 21.332 | 0 |
| 08:32:47.292 | -20.9606 | 170.2124 | 0 | -9154 | -9.154 | 0 |
| 08:32:47.245 | -21.05355 | 170.24808 | 0 | -30171 | -30.171 | 0 |
| 08:32:47.267 | -21.17935 | 169.96044 | 0 | -22839 | -22.839 | 0 |
| 08:32:51.392 | -27.47199 | 164.25869 | 1 | -11250 | -11.25 | 14966 |
| 08:32:51.413 | -27.44847 | 164.55369 | 1 | -3824 | -3.824 | 14095 |
| 08:32:52.954 | -28.16133 | 163.38961 | 1 | -9750 | -9.75 | 19320 |
| **08:32:54.700** | **-28.3123** | **162.971** | **0** | **128727** | **128.727** | **0** |
| 08:32:52.952 | -28.17513 | 163.35322 | 1 | -8236 | -8.236 | 13796 |
| 08:32:55.084 | -28.28163 | 163.2645 | 0 | -10405 | -10.405 | 0 |
| 08:32:54.741 | -28.33185 | 163.15856 | 1 | 4532 | 4.532 | 12085 |
| 08:32:54.933 | -28.28125 | 163.57713 | 0 | -4412 | -4.412 | 0 |
| 08:32:55.040 | -28.32231 | 163.45128 | 1 | -3485 | -3.485 | 14112 |
| 08:32:55.365 | -28.46071 | 163.33067 | 1 | 4129 | 4.129 | 18920 |
| 08:32:58.115 | -27.59998 | 164.39785 | 1 | -11860 | -11.86 | 18884 |